\documentclass[floatfix,aps,pra,twocolumn,showpacs,superscriptaddress,10pt]{revtex4-2}

\usepackage[utf8]{inputenc}
\usepackage[english]{babel}
\usepackage{subcaption}
\usepackage{mathtools}
\usepackage{float}
\usepackage{siunitx}
\usepackage{amsmath}
\usepackage{amsfonts}
\usepackage{amssymb}
\usepackage{graphicx}
\usepackage{color}
\usepackage{xcolor}
\usepackage[colorlinks=true,citecolor=blue,linkcolor=blue,urlcolor=blue]{hyperref}
\usepackage{times}
\usepackage{amstext}
\usepackage{latexsym}
\usepackage[running,mathlines]{lineno}
\usepackage{verbatim}
\usepackage{physics}
\usepackage{bm}
\usepackage{qcircuit}
\usepackage[version=3]{mhchem}
\usepackage{lipsum}  
\usepackage{ragged2e}
\newcommand{\beq}{\begin{equation}}
\newcommand{\eeq}{\end{equation}}
\newcommand{\barr}{\begin{eqnarray}}
\newcommand{\earr}{\end{eqnarray}}

\begin{document}
\raggedbottom

\title{Warehouse optimization using a trapped-ion quantum processor}
%Discuss title

\author{Alexandre C. Ricardo}
\affiliation{Departamento de F\'{i}sica, Universidade Federal de S\~{a}o Carlos, 13565-905 S\~{a}o Carlos, S\~{a}o Paulo, Brazil}

\author{Gabriel P. L. M. Fernandes}
\affiliation{Departamento de F\'{i}sica, Universidade Federal de S\~{a}o Carlos, 13565-905 S\~{a}o Carlos, S\~{a}o Paulo, Brazil}

\author{Amanda G. Valério}
\affiliation{Departamento de F\'{i}sica, Universidade Federal de S\~{a}o Carlos, 13565-905 S\~{a}o Carlos, S\~{a}o Paulo, Brazil}

\author{Tiago de S. Farias}
\affiliation{Departamento de F\'{i}sica, Universidade Federal de S\~{a}o Carlos, 13565-905 S\~{a}o Carlos, S\~{a}o Paulo, Brazil}

\author{Matheus da S. Fonseca}
\affiliation{Departamento de F\'{i}sica, Universidade Federal de S\~{a}o Carlos, 13565-905 S\~{a}o Carlos, S\~{a}o Paulo, Brazil}

\author{Nicolás A. C. Carpio}
\affiliation{Departamento de F\'{i}sica, Universidade Federal de S\~{a}o Carlos, 13565-905 S\~{a}o Carlos, S\~{a}o Paulo, Brazil}

\author{Paulo C. C. Bezerra}
\address{Wernher von Braun Advanced Research Center}

\author{Christine Maier}
\affiliation{Alpine Quantum Technologies GmbH, 6020 Innsbruck, Austria}

\author{Juris Ulmanis}
\affiliation{Alpine Quantum Technologies GmbH, 6020 Innsbruck, Austria}

\author{Thomas Monz}
\affiliation{Alpine Quantum Technologies GmbH, 6020 Innsbruck, Austria}

\author{Celso J. Villas-Boas}
\affiliation{Departamento de F\'{i}sica, Universidade Federal de S\~{a}o Carlos, 13565-905 S\~{a}o Carlos, S\~{a}o Paulo, Brazil}

\begin{abstract} 
Warehouse optimization stands as a critical component for enhancing operational efficiency within the industrial sector. By strategically streamlining warehouse operations, organizations can achieve significant reductions in logistical costs such as the necessary footprint or traveled path, and markedly improve overall workflow efficiency including retrieval times or storage time. Despite the availability of numerous algorithms designed to identify optimal solutions for such optimization challenges, certain scenarios demand computational resources that exceed the capacities of conventional computing systems. In this context, we adapt a formulation of a warehouse optimization problem specifically tailored as a binary optimization problem and implement it in a trapped-ion quantum computer.  

\end{abstract}

\maketitle

\section{Introduction}

Warehouse organization is essential to ensure the efficient operation of modern industries, being the major link between production and distribution channels. One of the most relevant tasks addressed by the management of a warehouse is the optimization of inventory control, which ensures that businesses can meet customer demands while minimizing surplus stock. An efficient warehouse management system aims to reduce the risk of stockouts or overstock situations while minimizing the time spent organizing the warehouse. 

In this context, determining the optimal placement of items on warehouse shelves, considering the layout and all possible combinations, often leads to an exponential increase in configurations as the number of products grows. Such problems, which involve identifying the optimal solution from a finite set of elements, are known as combinatorial problems \cite{schrijver2003combinatorial}. 

Combinatorial problems represent a broad category of mathematical challenges that are concerned with the organization, analysis, and optimization of discrete elements. These problems exhibit a wide range of complexities, varying significantly from one task to another. For instance, examples of these problems include partitioning problems \cite{Fan_Pardalos_2010}, the traveling salesman problem \cite{Laporte1992,Punnen2007}, and the knapsack problem \cite{Kellerer2004, Cacchiani2022_1, Cacchiani2022_2}. 

%The solution space of combinatorial problems often grows exponentially. Consequently, as the size of the problem increases, the search process becomes increasingly challenging due to the vast number of potential outcomes.
Due to the importance of these tasks, specialized computational methods have been developed to tackle combinatorial problems, among which simulated annealing \cite{press2007numerical} and genetic algorithms \cite{Potvin1996-qz} are notable. Over the last decade, the efforts to solve such problems using quantum computing have increased with the developments of hybrid quantum algorithms such as the Quantum Approximate Optimization Algorithm (QAOA) \cite{farhi2014quantum}, Variational Quantum Eigensolver \cite{Peruzzo2014-ji}, Digitized-Counterdiabatic Quantum Approximate Optimization Algorithm (DC-QAOA) \cite{PhysRevResearch.4.L042030}, and Quantum Annealing (QA) \cite{Mukherjee2015}. Advances in these algorithms have intensified interest among researchers in solving combinatorial problems, with recent evidence suggesting that, for specific cases, modified versions of QAOA can achieve up to exponential speedups compared to classical heuristics. \cite{zhang2024groverqaoa3satquadraticspeedup,doi:10.1126/sciadv.adm6761,Golden_2023,boulebnane2022solvingbooleansatisfiabilityproblems}.

In this work, an instance of the warehouse management problem developed in \cite{GabrielMatheus2024} is formulated as a Quadratic Unconstrained Binary Optimization (QUBO) problem, and solved using the QAOA in simulators of quantum hardware and a real trapped-ion quantum processor. This work is structured as follows: In Sec.~\ref{Sec:TF} we present a simple construction of the warehouse problem as a QUBO problem and an introduction to QAOA. In Sec.~\ref{Sec:Results}, we present the results of simulations of the algorithm in different scenarios and analyze its efficiency. Additionally, we demonstrate its practical applicability by executing our methodology on a trapped-ion quantum computer. In Sec.~\ref{Sec:Conclusions} we present our conclusions.

\section{Warehouse allocation and modeling problem}\label{Sec:TF}

Understanding the diversity of warehouse models is important when formulating strategies to optimize inventory management within a factory setting. An effective approach ought to take into account specific attributes, such as the racking type utilized in the production facility and dedicated shelves for specific products. Numerous studies are dedicated to developing methods that manage specific inventory models in factories such as multi-layered racks, similar to those commonly found in retail environments \cite{nastasi2018implementation, satori2023quantum, wang2016storage, lesch2021case, huang2022benchmarking}. A common shelving system used in factory warehouses is known as gravity flow shelving \cite{Tompkins1998-qd}. 
This type of shelf is engineered to facilitate the automatic movement of pallets along slightly inclined tracks, enabling their displacement via gravitational force. Warehouses with gravitational shelving usually apply the \textit{First In, First Out} (FIFO) \cite{Sazvar2016-ne,Onal2015-gv} principle, meaning that items are retrieved in the order of their insertion. In industrial warehouses, both the quantity of shelves and the number of shelf positions can be extensive, often reaching into the thousands or tens of thousands. As a result, an ideal warehouse within this principle would have more shelves than distinct products, which is impractical over large production facilities.

In this section, we will provide a concise overview of the warehouse allocation problem, which involves determining the optimal assignment of items from multiple gravitational shelves to fulfill specific order flows while minimizing costs such as transportation and storage time, along with an introduction to the Quantum Approximate Optimization Algorithm that will be employed to solve it.

\subsection{The warehouse allocation problem}

\par
\begin{figure}   
\begin{subcaptiongroup}
\includegraphics[clip, width=0.95\columnwidth]{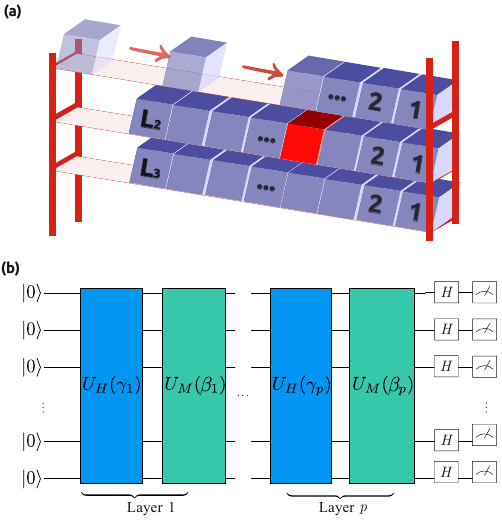}
    \phantomcaption \label{subfig-shelves}
    \phantomcaption \label{subfig-circuit}
\end{subcaptiongroup}
    \caption{\justifying Schematic representation of the problem: (\subref{subfig-shelves}) Gravitational racks that comprise the warehouse, filled with different products (boxes). Each shelf has a specific product capacity $L$ associated with it. In this example, if the production line requires only the red product, the other products in front of it should be removed from the shelf and then reinserted again. In this sense, managing the most efficient arrangement of products to be distributed through the shelves involves finding an organization that minimizes the number of reinsertions needed in the warehouse flow. (\subref{subfig-circuit}) The quantum algorithm applied in this work is a modified version of the QAOA. The parametrized quantum circuit is comprised of alternating layers of the evolution operator of the problem Hamiltonian, $\hat{H}$, and the time evolution operator of the mixer Hamiltonian, $\hat{H}_M$. The optimization parameters $\{\boldsymbol{\gamma},\boldsymbol{\beta}\}$ are indicated by a classical optimization subroutine.}
    \label{fig:fig1}
\end{figure}

 This study analyzes a Just in Sequence (JIS) inventory model that utilizes gravity racks, as schematically represented in Fig.~\ref{subfig-shelves}. Gravity racks are systems designed to move pallets along inclined tracks using gravitational force. This process involves assigning stock-keeping units (SKUs) and their respective SKUid codes (related to the type of packed object) to a specific shelf position upon arrival at the warehouse. However, this model can be challenging since, when requesting a particular SKU in the middle of the shelf, it is necessary to first remove the SKUs positioned in front of it.

Due to the sequential aspect of the JIS model applied to gravitational shelving, any additional items that are not requested but are collected must be restored in an available location. This process of reinsertion, while not ideal, can increase labor and machinery costs, as well as traffic and the likelihood of accidents and incidents. Typically, a warehouse does not need to be stocked with a large quantity of products simultaneously. The primary goal of this method is to strategically allocate a limited number of items within a spacious warehouse, featuring a higher number of distinct products than shelves, requiring that the frequency of item reinsertions must be minimized while storing different distinct products over shelves.

%\par To determine the optimal state of the warehouse, this study selected a configuration that minimizes transit and reinsertion. This task was completed by grouping items with dependencies or correlations together during insertion. This is made possible by finding the correlation matrix using a customized recommendation algorithm. Two different optimization approaches were tested using this matrix and the same data: QAOA, as detailed in this article, and annealing algorithms, as described in the article \cite{GabrielMatheus2024}.

Following the prior work modeling the warehouse problem as a QUBO Hamiltonian \cite{GabrielMatheus2024}, in this study we consider a warehouse comprised of $M$ gravitational shelves, each with a capacity limit $L_m$ associated with the shelf $m$, which could be a quantity of weight, number of items or space. At a given moment, this warehouse should be filled with $P$ new items, for each item $\alpha$ is assigned a weight $c_\alpha$. The units of measure of $c_\alpha$ and $L_m$ should agree since it is $L_m$ that defines how many items fit on the shelf $m$. In this work, it is considered only the case where $\sum_{\alpha} c_\alpha \leq \sum_{m} L_m$, i.e., the number of items to be allocated is always less than or equal to the number of available spaces. 

Another essential characteristic in this model is the interproduct cost matrix $\lambda$, with elements $0<\lambda_{\alpha\beta}<1$ that relates how costly is to store the pair of products $\{\alpha,\beta\}$ together. These parameters are essential when it is not feasible to have product-dedicated shelves—i.e., each shelf containing only one specific product. In the more common scenario, where different products must share the same shelf, a learning technique is employed to evaluate the affinity between each product pair. Solving the QUBO problem developed below with the parameters obtained via this technique ensures that items frequently taken out of the warehouse together have a low cost. In contrast, items rarely taken together incur a higher cost.

The formulation of the warehouse allocation problem used in this work can be divided into three aspects. First, every product must be allocated on a shelf. This condition imposes a constraint that can be represented by the penalty term $f_A$ in Eq.~\ref{Eq:fA}. Second, the total interproduct cost $f_B$ between items allocated on the same shelf must be minimized. This procedure guarantees that similar items are allocated in such a way that reduces the cost associated with storing different items together, and is represented by Eq.~\ref{Eq:fB}. Third, the capacity limit of each shelf must be respected, while similar items are allocated together. This constraint is represented by Eq.~\ref{Eq:fC} as cost $f_C$.

%The warehouse allocation problem can be translated into solving a QUBO problem composed by weighted sum of the following three main cost functions 
%
\begin{subequations}\label{Eq:Prob_ham}
\begin{align}
    f_A =  \sum_{\alpha=1}^P\left(1 - \sum_{m=1}^{M} x_{\alpha}^m\right)^2,\label{Eq:fA}\\
    f_B = \sum_{m=1}^{M}\left(\sum_{\alpha,\beta=1}^P \lambda_{\alpha \beta} x_{\alpha}^{m} x_{\beta}^{m}\right),\label{Eq:fB}\\
     f_C = \sum_{m=1}^M\sum_{\alpha=1}^P \left(c_{\alpha}  x_{\alpha}^{m} + \braket{\bm{2}}{\bm{a}_m} - L_m\right)^2.\label{Eq:fC}
\end{align}
\end{subequations}
The binary variables $x_\alpha^m$, represents if the product $\alpha$ is $(1)$ or is not $(0)$ placed in the shelf $m$. The term $\bra{\bm{2}}\ket{\bm{a}_m} = \sum_i 2^i x_{P+i}^{m}$ generates virtual products to fill the empty spaces of the shelves that should not be occupied in the optimal configuration. Therefore, the QUBO problem of minimizing the function $f=Af_A +Bf_B+Cf_C$, where $A, B$ and $C$ are weights for each function, is analogous to the quadratic problem of minimizing the function $f_B$ under the constraints attributed by $f_A$ and $f_C$ aforementioned. 

\subsection{The QAOA}

Variational quantum algorithms (VQAs) \cite{Cerezo_2021} have emerged as promising candidates for achieving quantum benefit on Noise Intermediary-Scale Quantum (NISQ) devices \cite{Preskill_2018}. They are composed of parameterized operations that are modified to optimize an object function. Their adaptability and the ability to operate without the need for complex error correction methods make them uniquely suited for current quantum computing capabilities. %Moreover, their structural resemblance to machine learning models places VQAs as an important development in the quantum machine learning research area. 
Among such algorithms, the Quantum Approximate Optimization Algorithm \cite{farhi2014quantum} stands out. Mostly known for its effectiveness in solving combinatorial and QUBO problems \cite{PunnenQUBO}, QAOA represents a hybrid approach. It combines a quantum subroutine, inspired by the adiabatic theorem, with classical optimization techniques.

In order to solve any combinatorial problem that can be encoded in a QUBO, the first step of the algorithm is to map the solution of the problem into the lowest energy state of a quantum operator $\hat{H}$, denominated \textit{problem Hamiltonian}. This mapping is achieved by translating binary variables $x_i$ into quantum operators with binary eigenvalues, thereby correlating the solutions of the problem with the eigenstates of these operators. After the translation mapping procedure, a quantum Hamiltonian $\hat{H}_M$, denominated \textit{mixer}, with known ground state, is chosen. To construct the circuit of the quantum algorithm, the qubits are prepared in the ground state of the mixer Hamiltonian, followed by alternating evolution operators of the problem Hamiltonian $\hat{H}$ and the mixer Hamiltonian $\hat{H}_M$, to simulate a Trotter-Suzuki decomposition \cite{Suzuki1976,Suzuki1977} of the adiabatic evolution in gate-based quantum computing, as shown in the circuit of Fig.~\ref{subfig-circuit}. In this algorithm, the time values in the evolution operators are treated as optimizable parameters that minimize the mean energy of the problem Hamiltonian. The probability of finding the solution to the problem increases with the number of alternating evolution operators, defined as the number of layers $p$.  However, the optimal depth of the quantum circuit required to minimize a cost function may have complexity QCMA-Hard (which can be seen as a quantum analog of the classical NP problems) in the worst cases \cite{bittel_et_al:LIPIcs.CCC.2023.34}.

%Since the QAOA was introduced in 2014, different variants of it have been proposed in order to improve specific aspects of the algorithm. Among these algorithms, Warm-Starting QAOA (WS-QAOA) \cite{Egger2021warmstartingquantum} aims to modify the initial state and mixer Hamiltonian based on the solution of the continuous relaxation of the problem, which can be solved with classical optimizers. Recursive QAOA (R-QAOA) reduces the search space recursively and, for specific Hamiltonians, may outperform the standard QAOA. There are also studies on constrained quadratic problems, that can be mapped into QUBO problems by adding penalty terms \cite{RUAN202398} or can have its search space restricted by using special mixers \cite{a15060202}.

The formulation provided in Eqs.~\ref{Eq:Prob_ham} is very general and can be applied to arbitrary number of objects to be allocated and arbitrary number of positions on the shelves. However, for simplicity and due to the limitations of the quantum hardware, in this work the standard version of the QAOA is applied to the case in which we want to allocate $P=3$ products over $M=2$ empty shelves, with capacity $L=2$ each. Since $L$ have the dimension of the number of items, the weight $c_\alpha$ is unitary. The interproduct cost matrix to be considered is:
\begin{equation*}
    \lambda = \begin{pmatrix}
     0.0 & 0.4 & 0.2\\
     0.4 & 0.0 & 0.6\\ 
     0.2 & 0.6 & 0.0
    \end{pmatrix}.
\end{equation*}
\noindent The values were chosen in a way that it is clear that the products $\alpha=3$ and $\alpha=1$ are less costly to be allocated together than with the product 2. The algorithm's current implementation can be divided into four steps: building of the Hamiltonian formulation, construction of the quantum circuit, evaluation of the mean energy function, and classical optimization.

\textbf{Building of the Hamiltonian formulation}: While different Hamiltonians can correctly represent the cost function, the circuits generated by the time evolution of these Hamiltonians may become hard to simulate using native gates available in real devices. Since trapped-ion quantum computers have native support for the Mølmer–Sørensen (MS) gate \cite{PhysRevA.62.022311, PhysRevLett.82.1835, PhysRevLett.82.1971}, it is beneficial to consider mappings that incorporate this gate. Therefore, we use the following map
\begin{subequations}
\begin{equation}\label{work_qubits}
    x_{\alpha}^m \longrightarrow \frac{1}{2}\left(\mathbb{I} - \hat{\sigma}_{x}\right)_{(m+(\alpha-1) M)},
\end{equation}
\begin{equation}\label{aux_qubits}
    a_m^l \longrightarrow \frac{1}{2}\left(\mathbb{I}-\hat{\sigma}_x\right)_{(M(P+l)+m)},
\end{equation}
\end{subequations}
\noindent where the indexes represent the qubits in which the operators are acting on and $\hat{\sigma}_x$ is the Pauli-X operator.
By mapping the binary variables in this way, the problem Hamiltonian that encodes the cost function can be written as 
\begin{equation}\label{gen_Hamiltonian}
    \hat{H} = \sum_{i\neq j} J_{i,j}\, \hat{\sigma}_x^{i}\,\hat{\sigma}_x^j + \sum_{i} h_i\, \hat{\sigma}_x^i,
\end{equation}
\noindent where the parameters $J_{i,j}$ and $h_{i}$ come directly from the mapping procedure and depend on the parameters of the problem. It is important to observe that the time evolution operator $U_H(t)=e^{-i\hat{H}t}$ of the problem Hamiltonian is composed only by $\hat{R}_{XX}$ and $\hat{R}_X$ operators \cite{Maslov_2017, cross2017open}.

\textbf{Construction of the quantum circuit}: The mixer Hamiltonian should have a ground state easy to prepare and not commute with the problem Hamiltonian. Thus, the mixer Hamiltonian is chosen to be
\begin{equation}\label{mixer}
    \hat{H}_M = -\sum_i \hat{\sigma}_z^i,
\end{equation}

\noindent allowing us to run the quantum algorithm and obtain an approximate solution in the basis spanned by the eigenstates of the $\hat{\sigma}_x$ operator, using a quantum circuit with $n = M(P+1+\log_2 L)$ qubits. Once defined the evolution operator $U_{M}(t)=e^{-i\hat{H}_M\, t}$ of the mixer Hamiltonian and the previously defined evolution operator $U_H (t)$ of the problem Hamiltonian, it is possible to construct the quantum circuit depicted in Fig.~\ref{subfig-circuit}, 
which generates, starting the qubits in the ground state $\ket{0 \dots 0}$, the output state
\begin{equation}
    \ket{\boldsymbol{\gamma},\boldsymbol{\beta}} = U_H(\gamma_n)\, U_M(\beta_n)\,\dots U_H(\gamma_1)\,U_M(\beta_1)\ket{0\dots 0},
\end{equation}

\noindent where $\{\boldsymbol{\gamma},\boldsymbol{\beta}\}$ are a set of parameters used in the optimization procedure.

\textbf{Evaluation of the mean energy function}: The population distribution of the states in the $\{\ket{+},\ket{-}\}$ basis is measured and used to evaluate the expected energy function $E(\boldsymbol{\gamma},\boldsymbol{\beta}) = \bra{\boldsymbol{\gamma},\boldsymbol{\beta}}H \ket{\boldsymbol{\gamma},\boldsymbol{\beta}}$. The expected energy function inversely correlates with the population of the ground state. As the expected energy decreases, the likelihood of the ground state population increases.

\textbf{Classical optimization}: The parameters $\{\boldsymbol{\gamma},\boldsymbol{\beta}\}$ are initialized randomly according to a uniform distribution and then the circuit is executed, running iteratively. The parameters are updated in each iteration until a minimum value of the expected energy is achieved. The task of updating the parameters between each execution of the circuit is done by a classical optimizer. In this work, the Py-BOBYQA \cite{10.1145/3338517}, a gradient-free solver with attested behavior with multistart methods \cite{8916288}, is used to obtain optimal parameters $\{\boldsymbol{\gamma}_{\text{optimal}},\boldsymbol{\beta}_{\text{optimal}}\}$.

\section{Experimental setup}\label{Sec:ExpSetup}
As quantum hardware for the real device implementation, we employ the quantum computer system IBEX hosted by AQT \cite{main_aqt}. The IBEX system is based on trapped ions, it supports a universal gate set and is accessible via the AQT cloud service ARNICA. The native gate set comprises single-qubit gates with arbitrary rotation angles and axis $\theta, \phi$: $R(\theta, \phi)=\mathrm{exp} [ -\mathrm{i}\theta/2 [\sin(\phi)\hat{\sigma}_y^{i}+\cos(\phi)\hat{\sigma}_x^{i}] ]$ 
as well as the two-qubit Mølmer-Sørensen (MS) gate as entangling operation with arbitrary rotation angle $\theta$ that can be implemented between any qubit pair $i, j$: $R_\mathrm{XX}(\theta)=\mathrm{exp}[-i\theta \hat{\sigma}_x^{i}\hat{\sigma}_x^{j}]$. In this work we utilize a 10-qubit register with all-to-all connectivity. We use Qiskit’s transpiler \cite{Qiskit_web} to compile the quantum circuits’ gates into
the hardware’s native gate set and to optimize the circuit for the minimal number of required gates. 
The gate errors are well approximated by depolarizing noise acting on the addressed qubits. The error rates for the single-qubit
gates are $(3.4 \pm 1.0)\cdot 10^{-4}$ on average over all qubits in the register. The error rate for the two-qubit gate corresponds to $(1.3 \pm 0.3)\cdot 10^{-2}$ for all pairs. 
Typical gate times are 22~$\mu s$ for single-qubit gates and 200~$\mu s$ for two-qubit gates. The T$_2$ time is $(100 \pm 6)$~$ms$, resulting in a coherence$/$gate time ratio in the order of $10^2$ \cite{ibex_aqt}.

\section{Results and Discussions} \label{Sec:Results}

The present section investigates the performance of the QAOA across different simulation scenarios and an implementation on a real trapped-ion quantum computer. 
%
%In the final subsection, a specific instance of QAOA is executed on a trapped-ion quantum computer. 
The experimental setup enables a comparison of the algorithm's efficacy when implemented on a quantum platform against its simulated counterparts. The subsequent analysis and comparison of results obtained from both simulated and experimental executions offer important insights into the practical viability and performance discrepancies inherent in contemporary quantum computing architectures.

\subsection{Resource estimation}

Resource estimation is an important analysis for assessing the feasibility and efficiency of models and algorithms. By providing information on expected computational resources, whether on real hardware, classical systems, or quantum simulators, resource estimation facilitates resource allocation while allowing comparisons with classical approaches.

By setting the individual weights $c_\alpha=1$, single qubit operations only need to be performed over the first $MP$ qubits. The number of qubits scales linearly with number of shelves and products and logarithmically with shelf capacity. Additionally, the number of two-qubit gates scales polynomially with the number of shelves and products. The summary of the estimated resources to run the algorithm can be found in Table~\ref{Table1}. 
\begin{table}[h!]
\begin{tabular}{|l|l|}
\hline
Resource           & Asymptotic behavior    \\ \hline
Number of qubits   & $M(P+1+\log_2 L)$      \\ \hline
Single qubit gates & $O(pMP)$               \\ \hline
$R_{XX}$ gates          & $O(pMP(M+P+\log_2 L))$ \\ \hline
\end{tabular}
\caption{\justifying Resources required to solve the warehouse allocation problem using the QAOA with $p$ layers. $M$ is the number of gravitational shelves, $L$ is the capacity limit of each shelf (assumed the same for all, for simplicity), and $P$ is the number of items to be allocated.}
\label{Table1}
\end{table}

The simulation of the quantum algorithm plays an important role when analyzing the outputs of heuristic algorithms, indicating how the quantum algorithm should perform in a perfect quantum environment. As an example, let us consider how many resources the algorithm would require in a real situation. To allocate $P=15$ products in a typical warehouse with effective $M=100$ shelves, each with capacity limit $L=8$, a total of $n=1900$ qubits are needed. As a consequence, given the challenges associated with simulating large systems, the analysis realized in the following sections is confined to a more manageable scenario of distributing $P=3$ items across $M=2$ shelves, which requires $n=10$ qubits.

\subsection{Simulation results}

In this section, various simulations of the algorithm were conducted to explore its performance under diverse scenarios. The simulations were executed considering the noiseless case using the Qiskit package \cite{Qiskit}. Different scenarios were considered to assess the algorithm's robustness and efficacy across different initialization conditions. These simulations also enabled a comprehensive evaluation of the algorithm's behavior when increasing the number of layers on the quantum circuit, allowing one to analyze the improvement in the mean energy and limitations given by the classical optimization subroutine and in practical quantum computing settings.

\begin{figure}[h!]
    \centering
    \begin{subcaptiongroup}    \includegraphics[clip, width = \columnwidth]{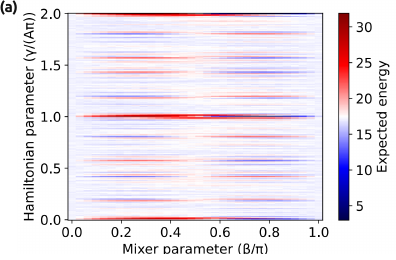}
    \includegraphics[clip, width = \columnwidth]{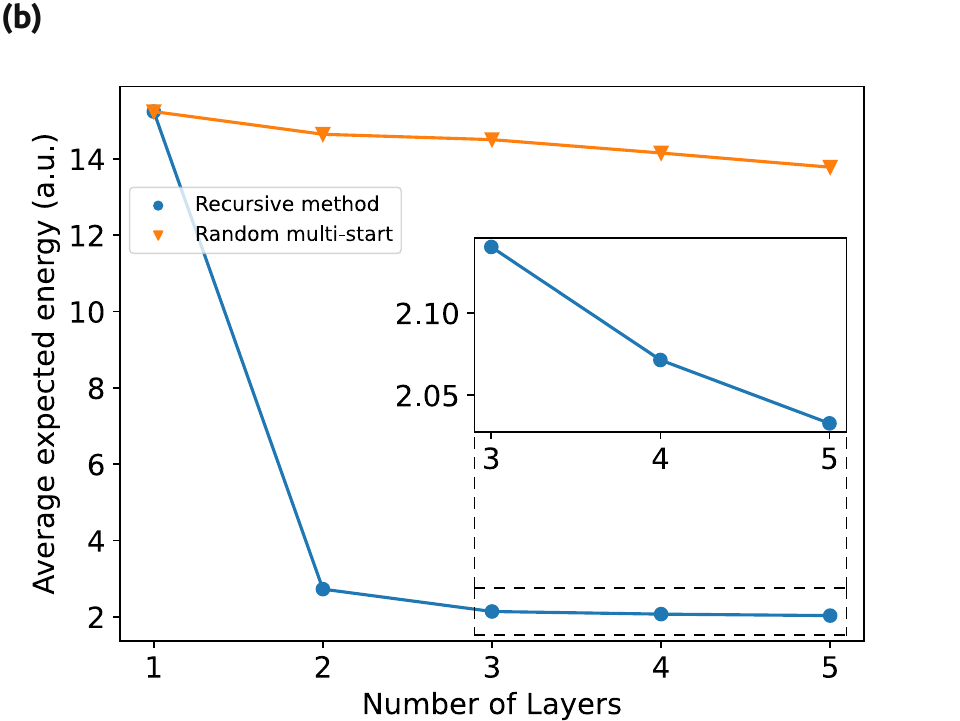}

    \phantomcaption \label{subfig-spectrum}
    \phantomcaption \label{subfig-methods}

    \end{subcaptiongroup}
    \caption{\justifying (\subref{subfig-spectrum}) Energy landscape simulated for the quantum algorithm. The parameteters of the problem resulting in this energy landscape are  $P=3$ products to be distributed over $M=2$ shelves, each with a capacity $L=2$, considering parameters $A=10$, $B=A/20$ and $C=B/2$. Each axis represents an optimization parameter for the mixer and problem Hamiltonian in the quantum circuit. (\subref{subfig-methods}) Average of the expected optimal energy values obtained by the two different strategies across the number of layers. The recursive method demonstrates a  pronounced reduction in the mean expected energy when increasing the layer quantity, in contrast to the outcomes observed with the random multi-start strategy.}
    \label{Fig:sim}
\end{figure}

The first scenario considered is the single-layer QAOA, Fig.~\ref{subfig-spectrum} illustrates the energy landscape obtained in the simulation. During the execution of the algorithm, this is the function that should be subjected to optimization by the classical optimizer to identify its global minimum. The simulated outcomes of the quantum algorithm indicate that there is a probability of $5.60\%$ to find either of the two potential ground states of the problem Hamiltonian in the global minimum of the expected energy. To ensure that the penalty for having a non-physical solution, \textit{i.e.}, a solution that assigns the same item on two different shelves at the same time, is high and the algorithm tends to search for physical solutions, the parameter $A$ must be larger than $B$ and $C$. In addition to that, not exceeding the limit of objects is equally essential as finding the ideal configuration given by the parameter $B$. Since the penalty term $f_C$ grows faster than $f_B$, the value of the parameter $C$ must be smaller than $B$.

% \begin{figure}[h!]
%     \centering
%     \includegraphics[clip, width = \columnwidth]{Figures/spectrum2.pdf}
%     \caption{\justifying Energy landscape simulated for the quantum algorithm. The characteristics of the problem resulting in this energy landscape are of $P=3$ products to be distributed over $M=2$ shelves, each with a capacity $L=2$, considering parameters $A=10$, $B=A/20$ and $C=B/2$. Each axis represent an optimization parameter for the mixer and problem Hamiltonian in the quantum circuit.}
%     \label{circuit_spectrum}
% \end{figure}

The energy landscape presented in Fig.~\ref{subfig-spectrum} reveals narrow gaps, indicating regions where classical optimization algorithms could potentially become trapped in local minima. As a result, the starting point of the optimization process directly influences the final outcome of the algorithm \cite{Lee_2021, Sack_2021, PhysRevA.107.062404}. This leads to the necessity of a different strategy in order to mitigate this issue and to understand the average behavior of the algorithm.

Two different strategies were considered in the simulations. The first was a random multi-start approach, involving multiple runs of the algorithm with distinct starting points. The results of these runs, shown as triangular orange markers in Fig.~\ref{Fig:sim}(b), indicate that as the number of layers increases, the mean expected energy decreases linearly. The average expected energy obtained with a single layer is  $\bar{E} = 15.24$, in arbitrary units, while for $p=5$ layers $\bar{E} = 13.78$ was achieved.  Additionally, increasing the number of layers not only improves the average minimum energy but also makes it easier to find lower-energy states, resulting in a higher population of such states. Conversely, the more layers we include in the algorithm, the larger the parameter space to be optimized by the classical routine becomes, hence increasing the computational cost of the optimization process. 

In the second mitigation strategy, a recursive approach is applied,  which comprises of a $p$-layer circuit, where the parameters of the first $p-1$ layers are fixed based on a previously obtained execution of the algorithm. The average expected energies vary from $\bar{E} = 15.24$, for a single layer, to $\bar{E} = 2.03$, for $p=5$ layers, as shown by the circular blue markers in Fig.~\ref{subfig-methods}. The expressive decrease rate of the first layers on the recursive approach is achieved due to the fact that it consists of the best results of previous layers. Moreover, it provides a systematic estimation of the solution to the problem. %Fig.~\ref{Fig:sim} shows the average expected energy obtained after optimization as a function of the number of layers. 
The average expected energy was obtained by averaging the optimal outcomes of $500$ executions of the algorithm with random starting points.

In a particular single execution of the algorithm, a random starting point is chosen for the parameters $\{\gamma,\beta\}$, within a single layer. Through iterative adjustments guided by the optimization subroutine, the single execution of the single-layer version of the algorithm aims to converge towards a stable, local minimum, configuration of the system. The algorithm yielded a $1.40\%$ probability of identifying the possible ground states, indicative of its capability to explore the system's energy landscape. It is important to acknowledge that this outcome does not represent the optimal performance of the algorithm, considered to be the global minimum of the expected energy landscape. The algorithm's performance can be influenced by factors such as the choice of optimization approach and the complexity of the energy landscape. The refinements or multiple executions presented in this section may be required to enhance the algorithm's efficacy in reaching optimal performance, especially when considering low-depth approaches.

\subsection{Real device implementation}

In this section, we detail the implementation of a specific instance of the algorithm on a trapped-ion quantum processor, with the main objective of validating the algorithm’s performance within an actual quantum computing
environment of a NISQ device. This practical execution is important for comparing the algorithm's theoretical predictions, derived from simulations discussed in the prior section, with its actual performance on quantum hardware. We seek to verify the algorithm's operation amidst the real-world conditions of quantum processors, characterized by noise and other imperfections. By comparing the outcomes from real-device experiments against simulated predictions, we investigate the discrepancies between theoretical expectations and the experimental results.

% \begin{figure}
%     \centering
%     \begin{subcaptiongroup}
%     \includegraphics[clip, trim = {0.5cm 0cm 1.5cm 0cm}, width=.95\columnwidth]{Figures/spectrum_AQT_200_inset.pdf}
%     %\includegraphics[clip, trim = {0.5cm 0cm 1.5cm 0cm}, width=.95\columnwidth]{Figures/spectrum_sim_28_inset.pdf}
%     \phantomcaption \label{subfig-hardware}
%      %\phantomcaption \label{subfig-sim}
%     \end{subcaptiongroup}
%     \caption{\justifying Complete energy spectrum of the problem hamiltonian described by Eqs.~\ref{Eq:Prob_ham}, considering a system composed of $P=3$ products to be inserted over $M=2$ shelves with capacity $L=2$ each, and weight parameters $A=10,$ $B=A/20$ and $C=B/2$. The blue dots represent every possible state and its respective energies. The black dots represent the global minima, and the red dots are the 10 most populated states obtained through the execution of the algorithm on AQT's trapped-ion quantum processor [...] system. These outputs represent different organization configurations of the warehouse.
%     }
%     \label{spectrum_fig}
% \end{figure}

\begin{figure}[h!]
    \centering
    \begin{subcaptiongroup}
    \includegraphics[clip, trim = {0.1cm 0cm 1.5cm 0cm}, width=.95\columnwidth]{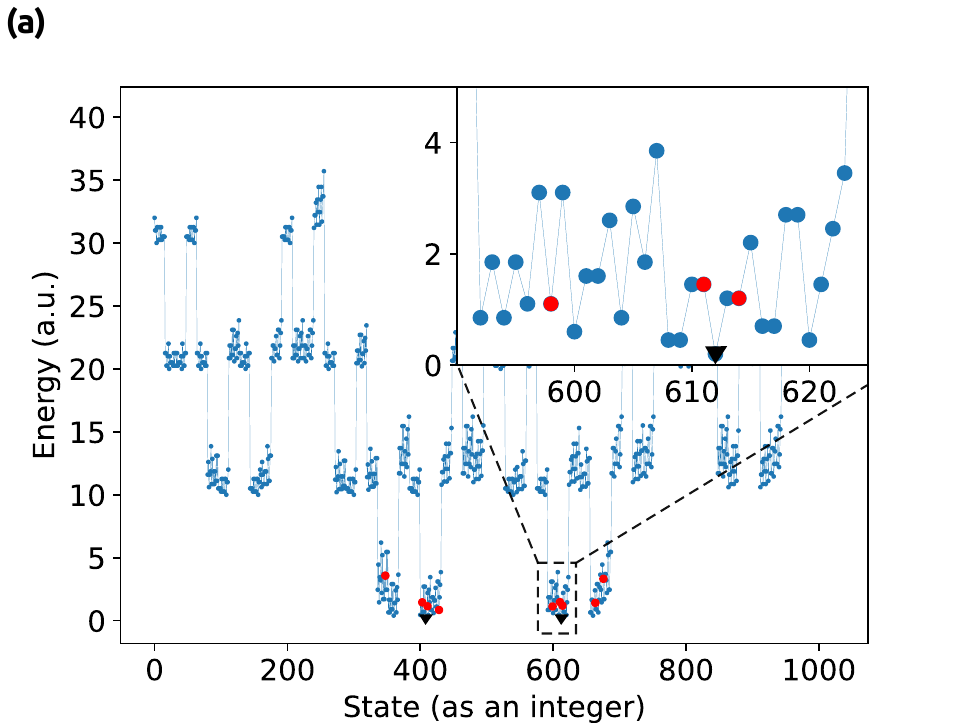}
    \includegraphics[clip, trim = {0.1cm 0cm 1.5cm 0cm}, width=.95\columnwidth]{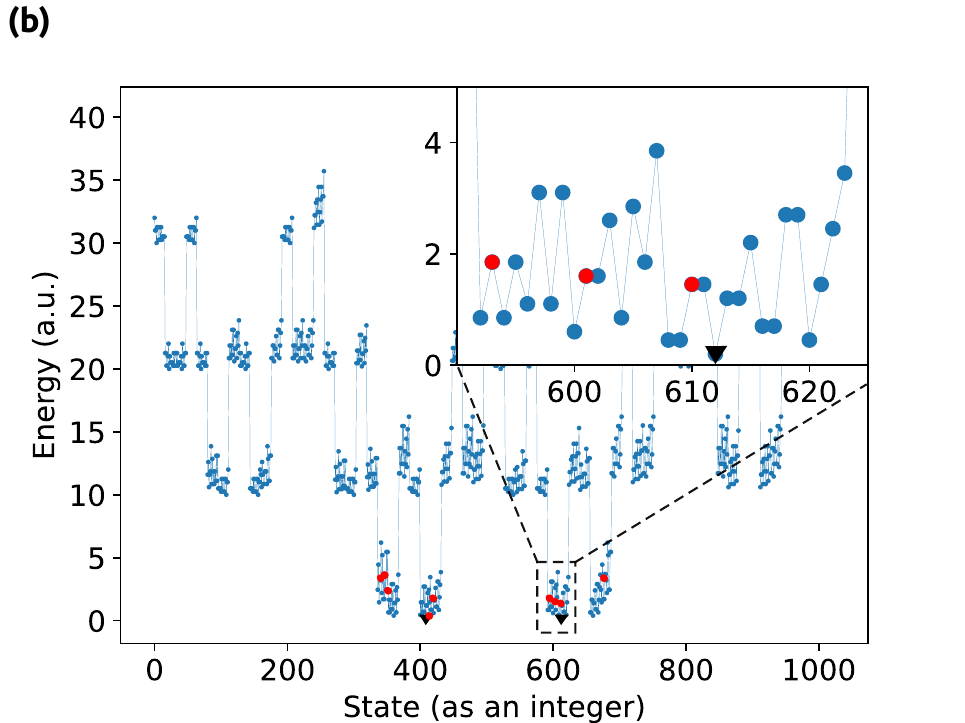}
    \phantomcaption \label{subfig-sim}
    \phantomcaption \label{subfig-hardware}
    \end{subcaptiongroup}
    \caption{\justifying Complete energy spectrum of the problem hamiltonian described by Eqs.~\ref{Eq:Prob_ham}, considering a system composed of $P=3$ products to be inserted over $M=2$ shelves with capacity $L=2$ each, and weight parameters $A=10,$ $B=A/20$ and $C=B/2$. The blue dots represent every possible state and its respective energies. The black triangles represent the global minima, and the red dots are the 10 most populated states obtained through (\subref{subfig-sim}) simulation and (\subref{subfig-hardware}) the execution of the algorithm on AQT's trapped-ion quantum processor system. These outputs represent different organization configurations of the warehouse. The plot shows agreement between the populations obtained through simulation and hardware. This agreement can also be quantified by comparing the final expected energy obtained, with $\hat{E}_{\text{sim}} = 2.40$ from simulation and $\hat{E}_{\text{hardware}} = 2.23$ from hardware execution.}
    \label{spectrum_fig}
\end{figure}

Fig.~\ref{spectrum_fig} illustrates the energy spectrum of the system for the current problem: Fig.~\ref{subfig-sim} shows the most populated states for the outcomes obtained after simulating the optimization process, while Fig.~\ref{subfig-hardware} refers to the optimization process executed on real hardware. Both simulation and hardware executions were performed with $n_{\text{shots}} = 200$ shots at each step. There is a notable similarity between the results; however, slight discrepancies are observed, which can be attributed to various factors inherent to quantum computing systems. For example, the populations and mean energies are slightly different between the simulation and the real hardware results, leading to different energy landscapes and, thus, different optimization paths. Other factors are the limitations imposed by the finite number of shots in the quantum circuit compared to the vast number of potential outcomes, and the presence of barren plateaus \cite{Wang2021,Sweke2020stochasticgradient,Xue_2021}.

In this particular execution, the expected optimal energy obtained was $\bar{E}_{\text{sim}} = 2.40$ and $\bar{E}_{\text{hardware}} = 2.23$ through simulation and hardware execution, respectively, further demonstrating the efficacy of the optimization process in minimizing the expected energy. A quantitative analysis demonstrating that the lower the expected energy obtained by this strategy, the more organized a warehouse is can be found in \cite{gabriel_patent}. These findings evidenced the potential of this quantum computing technique in accurately simulating and optimizing the distribution of products in a warehouse.

\section{Conclusions} \label{Sec:Conclusions}

In this study, we investigated a QAOA applied to a warehouse optimization problem. Through the examination of different initialization strategies and the exploration of increasing depth of the circuit, the robustness of the algorithm and its efficacy were evaluated. The simulations, particularly focusing on the single-layer QAOA setting, revealed promising outcomes with a global maximum probability of $5.60\%$ for identifying potential ground states in the particular scenario analyzed. 

An energy landscape analysis highlighted the challenges posed by local minima in classical optimization processes. A multi-start execution was performed to obtain the average behavior of the algorithm. This approach showed a linear decrease in the expected energy as a function of the number of layers, and did not converge to the global minimum within the used number of circuit layers. In order to improve the convergence, a recursive method was applied that explored the initialization sensitivity of parameters from the algorithm. Such a recursive approach improved the performance of the algorithm in the simulations, displaying a steep decrease in the average energy close to the global minimum.  Further refinements, such as selecting different optimization algorithms and optimizing the parameters $A, B$ and $C$, may be helpful in achieving optimal solutions, particularly in low-depth approaches.

The experimental validation of the optimization algorithm on the trapped-ion quantum processor provided proof of suitability for execution on existing devices. By executing the algorithm under realistic conditions, it was shown that the performance on a NISQ device reached similar results, measured by the expected minimal energy. The results showed that the implemented QAOA algorithm identified the optimal warehouse configuration in both simulation and execution on real quantum computing hardware.

\begin{acknowledgments}
This work was supported by the Coordenação de Aperfeiçoamento de Pessoal de  Nível Superior (CAPES) - Finance Code 001 and São Paulo Research Foundation (FAPESP) grants No. 2022/00209-6, 2023/14831-3, 2023/18240-0 and 2023/15739-3, the Austrian Research Promotion Agency (FFG) grant no. 884471 (ELQO) and the European Innovation Council (EIC) grant no. 190118992 (QCDC). C.J.V.-B. is also grateful for the support from the National Council for Scientific and Technological Development (CNPq) Grants No. 465469/2014-0 and 311612/2021-0. This work is also part of the MAI/DAI--CNPq Grants No. 139701/2023-0 and 141909/2023-4, CNPq 131088/2022-0 and CNPq 140467/2022-0.
\end{acknowledgments}

\bibliography{main}
%\bibliography{refs1}

\end{document}